\documentclass[10pt,a4paper]{article}
\usepackage{graphpap,epsfig,amssymb,graphicx,textcomp}
\usepackage[utf8]{inputenc}
\usepackage[english]{babel}
\begin{document}
\begin{center}{\Large{\bf Blume-Capel ferromagnet driven by 
propagating and standing magnetic field wave:
Dynamical modes and nonequilibrium phase transition}}
\end{center}

\vskip 1cm

\begin{center}{\it Muktish Acharyya$^1$ and Ajay Halder$^2$}\\
{\it Department of Physics, Presidency University}\\
{\it 86/1 College street, Calcutta-700073, India}\\
{E-mail(1):muktish.physics@presiuniv.ac.in}\\
{E-mail(2):ajay.rs@presiuniv.ac.in}\end{center}

\vskip 2cm

\noindent {\bf Abstract:}
The dynamical responses of Blume-Capel ($S=1$) 
ferromagnet to the plane propagating 
(with fixed frequency and wavelength) and standing magnetic field 
waves are studied separately in
two dimensions by extensive Monte Carlo simulation. 
Depending on the values of temperature, amplitude of
the propagating magnetic field and the strength of anisotropy, two different dynamical phases are observed.
For a fixed value of anisotropy and the amplitude of the propagating magnetic field, the system
undergoes a dynamical phase transition from a driven spin wave 
propagating phase to a pinned or spin frozen state as the
system is cooled down. The time averaged magnetisation over a full cycle of the propagating magnetic field plays the role of the dynamic order parameter.
A comprehensive phase diagram is plotted in the plane formed by the amplitude
of the propagating wave and the temperature of the system. It is found that the phase boundary shrinks
inward as the anisotropy increases. The phase boundary, in the plane described
by the strength of the anisotropy and temperature, is also drawn. This phase
boundary was observed to shrink inward as the field amplitude increases.

\vskip 1cm


\newpage

\noindent {\bf I. Introduction:}

Starting from its historical introduction, to analyse the thermodynamic 
behaviours of $\lambda$ - transition
in the mixture of He$^3$ - He$^4$, the Blume-Capel (BC) model\cite{blume,capel,emery}
became an important model, mainly to study the bicritical/tricritical behaviours
in various phase transitions. The low and high temperature series extrapolation
techniques were employed\cite{stauffer} to determine the different (continuous and
discontinuous) natures of the transition and hence the tricritical point (TCP) was
found in the fcc BC ferromagnet.
Monte Carlo study was performed\cite{jain} in fcc  BC model and traced a phase boundary
with tricritical point (TCP).
Later, Monte Carlo (MC) simulation within microcanonical ensemble
was studied\cite{deserno} in Blume-Capel model to show the tricriticality. 
The critical behaviours of $S={3 \over 2}$ was studied\cite{lara}.  
It should be mentioned here that the variety of positional and/or orientational
order and a very rich phase diagram was obtained\cite{yunus} 
recently even in $S={3 \over 2}$ Ising
system in three dimensions by renormalization group theory in Migdal-Kadanoff
approximation.
The general spin BC model was studied\cite{general} by meanfield approximation.
The meanfield solution was obtained\cite{santos} in BC model with random crystal
field. Effects of random crystal field in BC model was studied\cite{yuksel} by
effective field theory. 
The wetting transition in BC model was studied\cite{albano} by MC simulation.

Interestingly, the BC model exhibits the competing metastability which has an
important role in its dynamical behaviours. In this context, it should be mentioned
that dynamic Monte Carlo and numerical transfer matrix method \cite{fiig}
were employed to study the competing behaviours of the metastability in the BC model.
The behaviours of the competing metasatble states at infinite volume are studied  
\cite{manzo} in dynamic BC model. The metastable and unstable states are obtained
\cite{ekiz} by cluster variation and path probability method.

The researchers paid attention to study the
nonequilibrium behaviours (in the presence of time varying magnetic field)
 of Ising\cite{rev1}, Heisenberg\cite{rev2} and BC model. 
The nonequilibrium dynamic
phase transition was studied\cite{deviren} in $S={5 \over 2}$ 
BC model in the presence of an oscillating (in time but 
uniform over space) by effective field
analysis and Glauber dynamic approach. The nonequilibrium dynamics was studied\cite{vatansever} in $S={3 \over 2}$
BC model with quenched random fields. The effective field theory was 
developed\cite{ertas} for kinetic BC model driven by oscillating magnetic field
to study the dynamic phase transition. The meanfield approach was employed
\cite{keskin} to
study the dynamic phase transition in Glauber kinetic spin-1 blume-Capel model
swept by oscillating magnetic field.

What kind of nonequilibrium responses and the transitions are expected if the
Blume-Capel ferromagnet is swept by propagating and standing magnetic wave ? The
above mentioned nonequilibrium responses are due to an oscillating magnetic field
which is {\it oscillating in time but uniform over the space}. The system will respond
differently if the externally applied field has spatio-temporal variation. The
nonequilibrium responses of the Ising ferromagnet were recently studied by applying
propagating magnetic field wave and various interesting {\it spin wave} motion
and nonequilibrium phase transitions are observed\cite{JMMM1,JMMM2,appb}. 
The dynamic phase transitions, in the Ising ferromagnet, are studied recently
\cite{ajay} in the presence of standing magnetic field wave.

In this 
paper, the nonequilibrium responses of Blume-Capel ferromagnet ($S=1$) are studied
in the presence of propagating and standing magnetic field wave by Monte Carlo
simulation. The paper is organized as follows: the model is described in the section
II, the numerical results are reported in section III and the paper ends with summary
in section IV.

\vskip 0.5cm

\noindent {\bf II. Model and simulation:}

The energy of the two dimensional Blume-Capel ferromagnet in the 
presence of magnetic field (having spatio-temporal variation) is

\begin{equation}
E(t) = -J\sum S_z(x,y,t)S_z(x',y',t) +D\sum (S_z(x,y,t))^2 -\sum h(x,y,t)S_z(x,y,t)
\end{equation}

\noindent where, $S_z(x,y,t)$ represents the z component of the Spin ($S=1$) at any position
($x,y$) in time $t$. The values of $S_z(x,y,t)$  
may be any one of -1,0 and +1. $J(>0)$ is the uniform (over the space) ferromagnetic interaction
strength. The first term represents the ferromagnetic nearest neighbour 
spin-spin interaction. The second term provides the contribution due to the
anisotropy ($D$).
The last term represents the
interaction between spin and the externally applied magnetic field
having spatio-temporal variation. $h(x,y,t)$ denotes the value of 
magnetic field at position
$(x,y)$ at time $t$. Two types of the forms of $h(x,y,t)$ are chosen here
e.g., propagating wave and standing wave. 
The propagating magnetic field wave is represented as

\begin{equation}
h(x,y,t)=H{\rm cos}(2\pi f t - 2\pi y/{\lambda}).
\end{equation}

\noindent The standing magnetic field wave may be represented as

\begin{equation}
h(x,y,t)=H{\rm sin}(2\pi f t){\rm sin}(2\pi y/{\lambda}).
\end{equation}

For simplicity, the magnetic field wave, considered here, 
is propagating in the $y$ direction. The $H$, $f$ and $\lambda$ represent the
amplitude, frequency and the wavelength of the  
magnetic field waves (both propagating and standing) respectively.
The periodic boundary conditions are applied in both directions 
of the square lattice to preserve the translational
invariances in the system.

\vskip 0.5cm

\noindent {\bf III. Results:}

\noindent {\it (i) For propagating wave}

In this study, a square lattice of size $L=100$ is considered. The system is 
driven by a propagating (moves in the $y$ direction) magnetic field wave having
fixed wavelength $\lambda=50$ and frequency $f=0.01$. The frequency and 
wavelength of the propagating magnetic field wave are kept constant throughout
the study. The system ($L=100$) contains two full waves for this choice of 
particular value of $\lambda=50$. 
Initially, a high temperature ($T=2.0$ here) 
random phase is considered where the
system is prepared with a random uniform distribution of the values (1, 0, and
-1) of $S_z(x,y,t=0)$. This is a high temperature paramagnetic phase.
Now the system is slowly (with decrement of temperature
equals to 0.02) cooled to achieve the nonequilibrium steady state at any fixed
temperature $T$. In this Monte Carlo simulation, the Metropolis single 
spin flip algorithm with parallel 
updating\cite{springer} rule is employed. 
The probability of spin flip is chosen as ${\rm Min}[e^{{-\delta E} 
\over {kT}},1]$, where, $\delta E$ is energy required for spin flip and $k$ is
the Boltzmann constant. The field amplitude ($H$) and the strength of anisotropy
($D$) are measured in the unit of $J$ and the temperature are measured in the
unit of $J/k$.
$L^2$ such updating of spins defines the unit of 
time i.e., Monte Carlo step per spin (MCSS) here. At any temperature, the 
system is allowed to pass through $10^5$ MCSS and initial (or transient) 
$5\times 10^4$ MCSS are discarded. The dynamical quantities are calculated
over $5\times 10^4$ MCSS. It may be noted here, that for frequency $f=0.01$ of
the propagating magnetic field wave, 100 MCSS is required to have a complete
cycle of the propagating magnetic field wave. 
So, in $5\times 10^4$ MCSS, 500 such cycles of the propagating magnetic
field are present. It is checked that this length of simulation is adequate to
achieve the nonequilibrium steady state. The instantaneous magnetisation is 
$M(t)={{1} \over {L^2}} \sum_i S_z^i(x,y,t)$, the time averaged magnetisation
over the full cycle of the propagating magnetic field is 
$Q=f\times \oint M(t)dt$. Due to the intrinsic 
stochasticity in the Metropolis dynamics,
the values of $Q$ are different in different cycles. As a result, $Q$ has 
a statistical distribution. The variance of $Q$ is $V=L^2(<Q^2> - <Q>^2)$, 
where $<Q^2>$ is
average of $Q^2$ over 500 different values of $Q$. It is checked that this
number of samples is sufficient to have good statistics.

It is found that, for fixed values of $D$ and $H$, two different dynamical
phases are identified. In the high temperature, two distinct and
alternate bands of spin
values $S_z=+1$ and $S_z=-1$ are formed. They are found to propagate.
A few sites having $S_z=0$ are found along the boundary of the bands.
For sufficiently high values of the temperature ($T$) of the system 
and the amplitude ($H$) of the propagating magnetic field,
these spin-bands are observed to be propagating coherently 
along the direction of
propagation of the magnetic field wave. One such coherent propagation of
{\it spin-wave} is shown in Figure-1. 
It should be mentioned that the
term {\it spin-wave} used here is strictly 
different from the conventional notion
of spin-wave in condensed matter physics.

The low temperature dynamical phase is not the coherent propagation of
{\it spin-waves}. For fixed values of $D$ and $H$ as the system is cooled
down, the system transits to a different phase. This is a {\it spin-frozen}
or pinned phase, where most of the spins are pinned or frozen
to any one value of $S_z=\pm 1$.
Here, the propagating magnetic field wave does not have any effect on the
spins of the system. The spin flip is almost stopped in this phase. The two
major dynamical phases are observed, namely, the high temperature
{\it spin-wave} propagating phase and low temperature {\it spin-frozen} or 
pinned phase. In the {\it spin-wave} propagating phase, the populations
of $S_z=0$ depends on the value of the strength of anisotropy $D$. For lower
values of $D$, this population increases. The dependences of these dynamical
phases on $D$, $H$ and $T$ are shown in a comprehensive manner in Figure-2.
It may be noted here that the dynamical phase transition was studied
\cite{appb} also
in Ising ferromagnet swept by propagating magnetic field. A similar coherent
propagation of spin bands was observed there. However, in the Blume-Capel 
ferromagnet, in the low temperature ($T$) and high field ($H$) propagating
phase, the large population of $S_z=0$ was observed in the interfacial region
of spin bands formed by $S_z=+1$ and $S_z=-1$ (see Fig-2(iii)).

To study the dynamical phase transition quantitatively, the dynamical order
parameter $Q$ and its variance $V$ are studied as a function of temperature
with $H$ and $D$ as parameters. For fixed value of $D$ and $H$, in the high
temperature {\it spin-wave} propagating phase, the dynamical order parameter
$Q$ assumes zero value and in the low temperature region, $Q$ becomes nonzero.
Actually, modulus of $Q$ is greater than zero in the low temperature
{\it spin-frozen} phase. As the system is cooled down it undergoes a dynamical
phase transition from a {\it spin-wave} propagating phase to a 
{\it spin-frozen} phase. The order parameter $Q$ starts to assume 
a nonzero value at
any finite temperature $T$ which gives rise to a dynamical phase transition.
From the temperature dependence of $Q$, the transition seems to be continuous.
This dynamical phase transition is also detected from the temperature variation
of $V$, which shows a very sharp peak at the transition temperature. 
It is believed to diverge eventually in the $L \to \infty$ limit.
A typical such variation is shown in Figure-3. In this figure, the dependences
of the transition temperature on $D$, is also shown. For a fixed value of $H$,
the transition occurs at lower temperature as one increase the strength of 
anisotropy $D$. As an example, here for $D=1.0$ the peaks of $V$ are observed
at $T=0.40$ and $T=1.36$ for $H=1.5$ and $H=0.1$ respectively. For lower value
of anisotropy ($D=0.1$) the dynamic transitions occur at $T=0.70$ and $T=1.66$
for $H=1.5$ and $H=0.1$ respectively, indicated by the sharp peaks on $V$. 

This
dynamical phase transition detected jointly from the temperature dependences
of $Q$ and $V$ is found to be happened at lower temperature as one increase
the amplitude ($H$) of the propagating magnetic field wave keeping all other
parameters fixed. For any fixed value
of $D$ just by getting the transition temperature (from the peak position of
$V$) as a function of $H$, the comprehensive dynamical phase boundary
(in the plane formed by $H$ and $T$) is 
obtained and shown in Figure-4. The phase boundary is observed to shrink inward
(lower values of $T$ and $H$)
as the strength of anisotropy $D$ increases. The transitions observed, along the
entire phase boundary, are of continuous types. It may be noted here that the
equilibrium transition temperature (for $D=0$, $H=0$) is very close to 1.8 
(see Fig-2 of Ref\cite{albano}). 
The results for $D=0.1$, obtained here, may be compared to that.

The variances of $Q$ is studied as function of temperature for
 two different values of $D$ (with fixed $H$). This shows the transition occurs at lower temperature
for higher values of $D$. These are shown in Figure-5. As an example here, for
$H=1.0$ the transitions occur at $T=0.52$ and $T=0.76$ for $D=1.3$ and $D=0.8$
respectively. Similarly, for $H=0.5$ the transitions occur at $T=0.90$ and
$T=1.14$ for $D=1.3$ and $D=0.8$ respectively.

In a similar method, the dynamical phase boundaries are obtained and sketched
in the plane formed by $D$ and $T$ for two different values of $H$. It is shown
in Figure-6. It is observed that for large values of $H$, the phase boundary
shrinks inward (lower values of $T$ and $D$). 
Here also, along the entire phase boundary, the dynamical transitions
are found to be continuous.

\vskip 0.3cm

\noindent {\it (ii) For Standing wave}  

The dynamical phase transition was studied also in Blume-Capel ferromagnet 
irradiated by
standing magnetic wave. Here the frequency of the standing wave is 
chosen $f=0.01$ and the wavelength was taken $\lambda=20$. These choice of $f=0.01$
and $\lambda=20$ are kept fixed throughout the study.

Here also, like the case of propagating wave, two distinctly different dynamical modes
were observed. For a fixed set of values of $H$ and $D$, the low temperature phase
is a {\it spin-frozen} state and the high temperature phase shows the formation of
alternate ($S_z=1$ and $S_z=-1$) spin bands. These spin bands does not propagate.
The spin bands also form a {\it standing spin-waves}. To be precise, as the system is
cooled down from a high temperature, the system undergoes a nonequilibrium
phase transition from a {\it standing spin-wave} phase to a {\it spin frozen} phase.
These typical morphologies are shown in Figure-7.
It may be noted here, that similar standing spin bands were observed 
\cite{ajay} in the
case of Ising ferromagnet swept by standing magnetic wave. Here, no interfacial
population of $S_z=0$ was observed unlike
case of propagating wave (applied in BC model). 

To distinguish the {\it propagating spin wave} and the {\it standing spin wave} phases
the snap shots of the spin configurations of the lattice are shown for a fixed set
of values of $T$, $D$ and $H$ at two different times ($t=1000$MCSS and $t=1070$MCSS).
This is shown in Figure-8.
No signature of propagation of the alternate spin bands was observed. 
Reader may compare this with Figure-1, where the coherent propagation of spin bands
was observed with the application of external propagating magnetic field wave.

The temperature dependences of $Q$ and $V$ are studied using 
$H$ and $D$ as parameter and shown
in Figure-9. The sharp peaks of $V$ indicated the transition 
temperature of the nonequilibrium phase transition from high temperature
 {\it standing spin wave} to
low temperature {\it spin frozen} state. The transition temperatures
are found to depend on $D$ and $H$. This is comprehensively shown in the 
Table-I.

\begin{table}[ht]
\caption{Dependences of transition temperature ($T$) on H and D}
\centering
\begin{tabular}{|c| c| c|}
\hline\hline
H & D & $T_c$ \\
\hline\hline
0.50&0.10&1.50\\
\hline
0.50&0.50&1.38\\
\hline
1.00&0.10&1.20\\
\hline
1.00&0.50&1.10\\
\hline
\end{tabular} 
\end{table}
\vskip 0.5cm

The comprehensive phase boundaries in the H-T plane are plotted in different values of D and shown in
Figure-10. The phase boundary was found to shrink inward 
(lower values of $T$ and $H$)
as D increases. The transition was observed to be continuous through the entire phase
boundary.

The nonequilibrium phase transitions with application of standing magnetic wave are
studied in the plane described by D and T (for different values of H).
The comprehensive phase boundaries in the D-T plane are plotted and shown in 
Figure-11. Here also, the phase boundary was observed to shrink inward
(lower values of D and T) as H increases. The nature of the transition remains 
continuous through the entire phase boundary.

We have compared a phase diagram (for standing wave $D=0.5$ and $\lambda=20$)
of dynamic phase transition
with that obtained from a meanfield study
\cite{keskin} of Glauber kinetic $S=1$ BC model. The results are shown 
(Figure-12) in a plot
of suitably rescaled
temperature ($T'=T/z$) and field amplitude ($h'=H/z$). The variable $z$
represents the coordination number which is equal to 4 in the case of a square
lattice (in this study).

\vskip 1cm

\noindent {\bf IV. Summary:}

The dynamics of the Blume-Capel (S=1) ferromagnet 
in the presence of the plane propagating 
and standing
magnetic field waves are studied in
two dimensions by Monte Carlo simulation. The Metropolis single
spin flip algorithm with parallel spin updating rule are used. The different
dynamical states of the system are observed to depend on the values of
the strength of anisotropy, temperature of the system and the amplitude of
the plane magnetic field waves (both propagating and standing).
Here, mainly two different dynamical phases are observed.
For a fixed value of anisotropy and the amplitude of the 
propagating magnetic field, the system
undergoes a dynamical phase transition from a driven {\it propagating spin wave} 
phase to a pinned or spin frozen state as the
system is cooled down. In the driven spin wave state, the coherent motion
of the alternate spin-bands of $S_z=+1$ and $S_z=-1$ are found.

The time averaged magnetisation over a full cycle of 
the propagating magnetic field plays the role of the dynamic order parameter.
Accordingly, the spin frozen or pinned phase is characterised by nonzero value
of the dynamical order parameter. The dynamical order parameter becomes zero 
in the phase of coherent propagation of the spin-bands.  
This dynamical order parameter takes a nonzero 
value at the transition temperature as the
system is cooled down. The variance of the order parameter gets sharply peaked
at the transition point. The transition temperature is found from the peak
position of the variance of the dynamic order parameter. This dynamic transition
temperature was found to be function of amplitude of the propagating magnetic
field wave and the strength of the anisotropy. 
A comprehensive phase diagram is plotted in the plane formed by the amplitude
of the propagating wave and the temperature of the system. It is found that the phase boundary shrinks
inward (lower temperature) 
 as the strength of the anisotropy increases. 
The phase boundary, in the plane described
by the strength of anisotropy and temperature, is also drawn. Here, this phase
boundary was observed to shrink inward as the amplitude of the propagating 
magnetic field wave increases.

The dynamical responses of the Blume-Capel model was studied here with the application
of standing magnetic wave. Here, the high temperature phase is quite different from
that observed in the case of propagating magnetic wave. The standing wave 
(non propagating) of alternate
spin bands are formed. The dynamical phase boundaries are plotted in the H-T and D-T
planes. The qualitative behaviours are same as observed in the case of propagating
magnetic wave.
 
It may be noted that the sharp peak of the variance at the transition point
is a signature of the divergence (in the thermodynamic limit)
of the dynamic susceptibility. The Blume-Capel model shows tricritical 
behaviour in the equilibrium phase transition. However, in this 
particular nonequilibrium
case, the transition seems to be continuous irrespective of the value of
the strength of the anisotropy. 

A qualitative understanding of the existence of the dynamical phases may be
as follows: at low temperature the values D and H are inadequate for spin
flip. As a result one obtains the {\it spin frozen} phase. On the other hand
if the temperature is high the same set of values of D and H become strong enough
to flip the spins. In this phase spins are flexible enough to respond to the
variation of propagating and standing magnetic wave. As a result, the propagating
magnetic wave yields {\it propagating spin bands} and standing magnetic wave
produces {\it standing spin bands} in the lattice. Phase boundaries is nothing but
the dependences of the dynamic transition temperature on the values of D and H.

Recently, the site diluted BC model was studied\cite{expt1} 
by meanfield renormalization group
analysis with good agreement of the experimental phase diagram of Fe-Al alloy. 
The magnetic alloy of Fe-Al can be modelled
by Blume Capel ferromagnet and it would be interesting to see the 
coherent propagation of spin-bands and the standing spin band experimentally by 
time resolved magneto-optic Kerr (TRMOKE) effect. This study has a significance
in the field of spintronics and magnonics\cite{bader}

The magnetic behaviours of core-shell magnetic nanoparticles has an important
in the magnetism research as well as in the technology. The magnetic properties
have been studied\cite{zeng} in bimagnetic ($FePt/MFe_2O_4(M=Fe,Co)$) core-shell
nanoparticles. The dynamical phase transition has been studied
\cite{polat} by Monte Carlo
simulation in spherical core-shell ($S={3 \over 2}$ core and $S=1$ shell)
under time dependent (uniform over space) magnetic field. It would be 
interesting to study the dynamical behaviours of core-shell magnetic 
nanoparticles in presence of a magnetic field having spatio-temporal variation
appearing as magnetic propagating wave, standing wave etc.

\newpage

\begin{figure}[h]
\begin{center}
\begin{tabular}{c}
\resizebox{7cm}{!}{\includegraphics[angle=0]{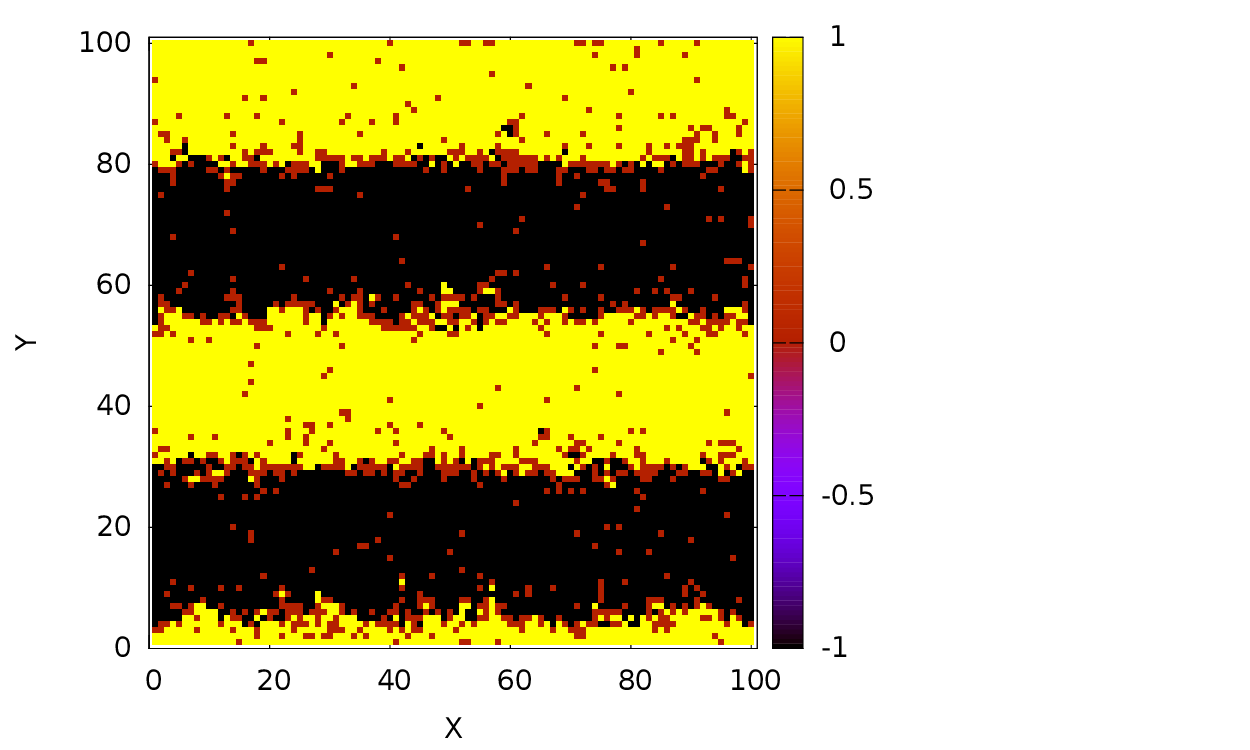}}
\\
\resizebox{7cm}{!}{\includegraphics[angle=0]{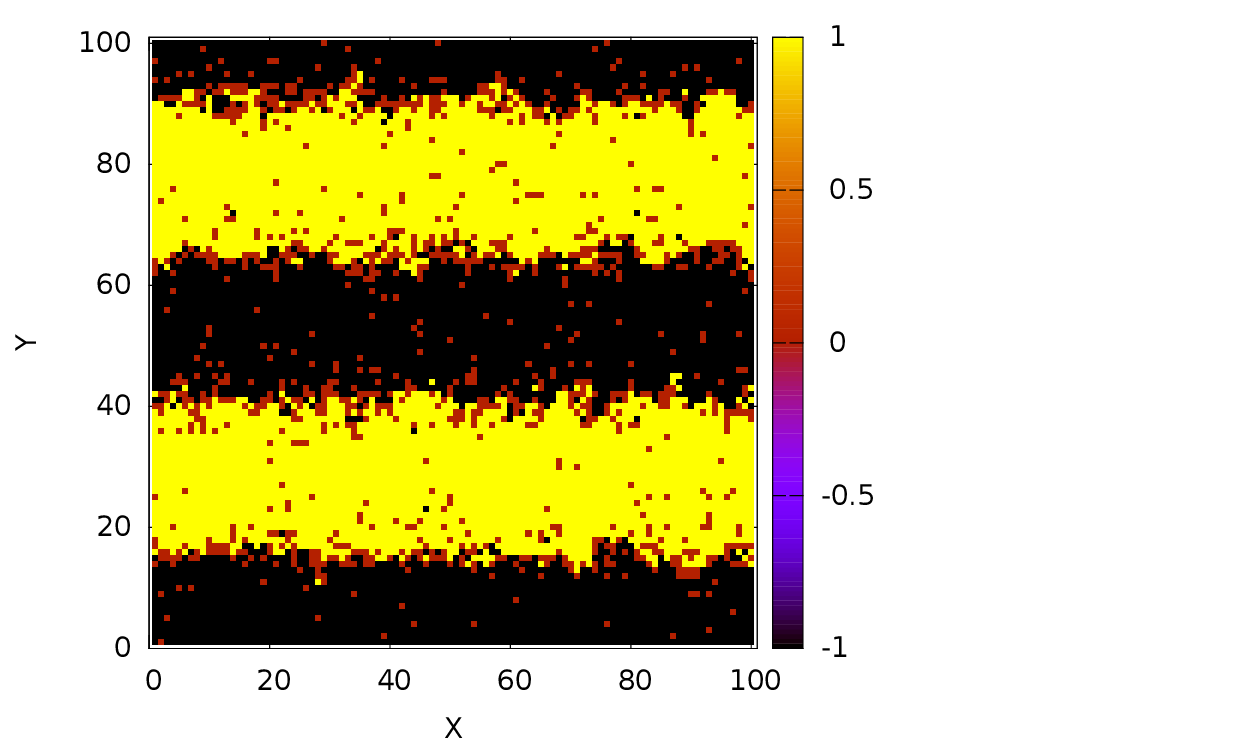}}
          \end{tabular}
\caption{Coherent propagation of driven spin-wave 
shown here for two different values 
of time. Top is for t=2000 and bottom is for t=2070. 
Here, D=1.0, H=1.5, T=1.0}

\end{center}
\end{figure}

\newpage
\begin{figure}[h]
\begin{center}
\begin{tabular}{c}
\resizebox{7cm}{!}{\includegraphics[angle=0]{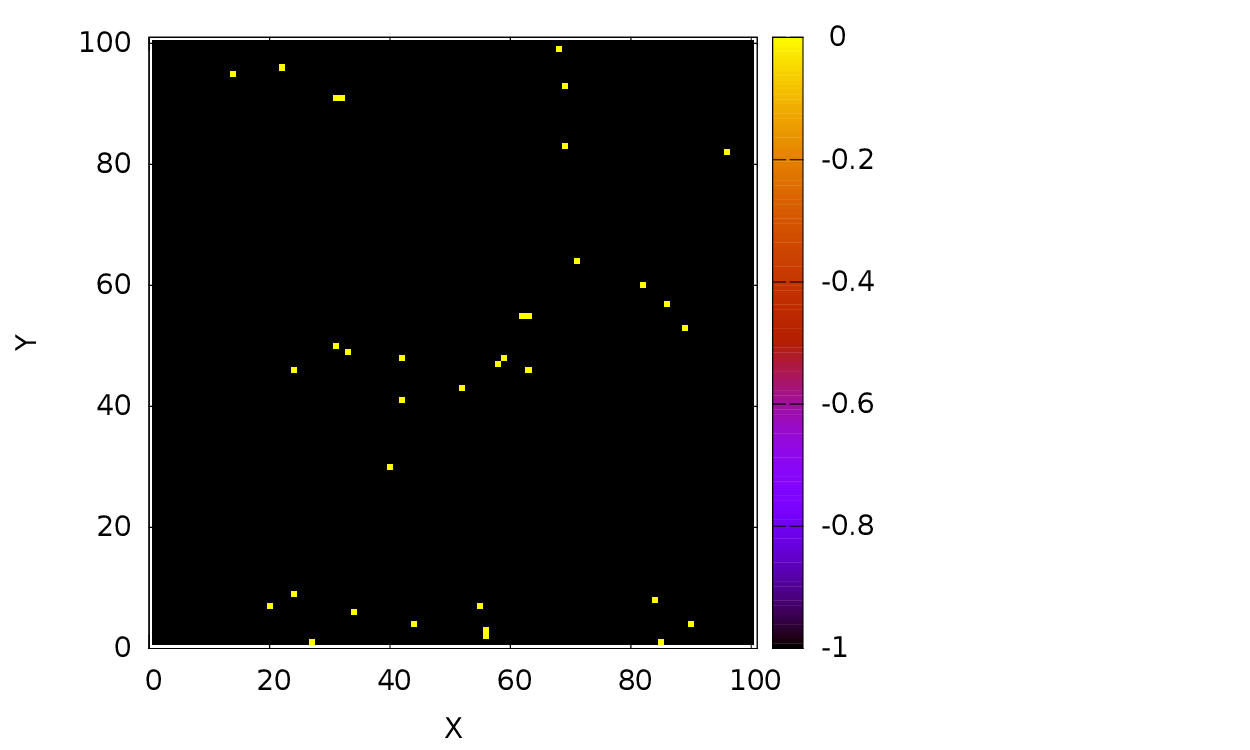}}
\resizebox{7cm}{!}{\includegraphics[angle=0]{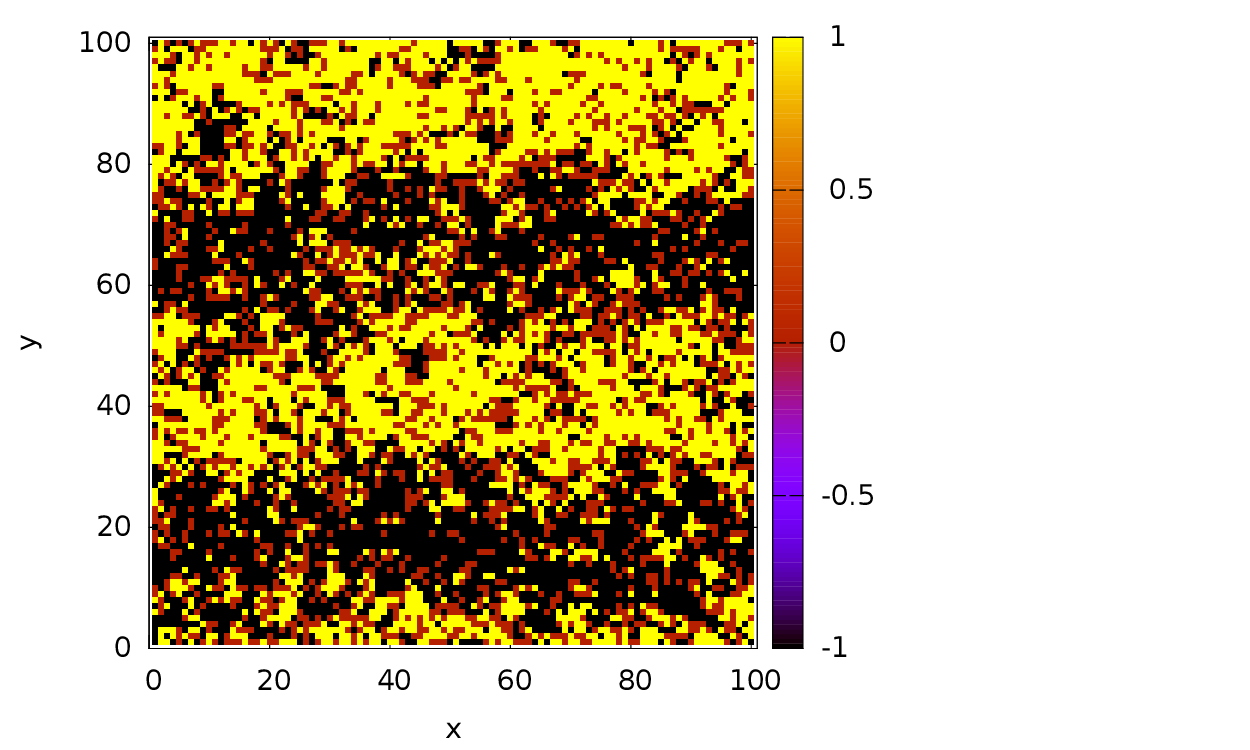}}
\\
\resizebox{7cm}{!}{\includegraphics[angle=0]{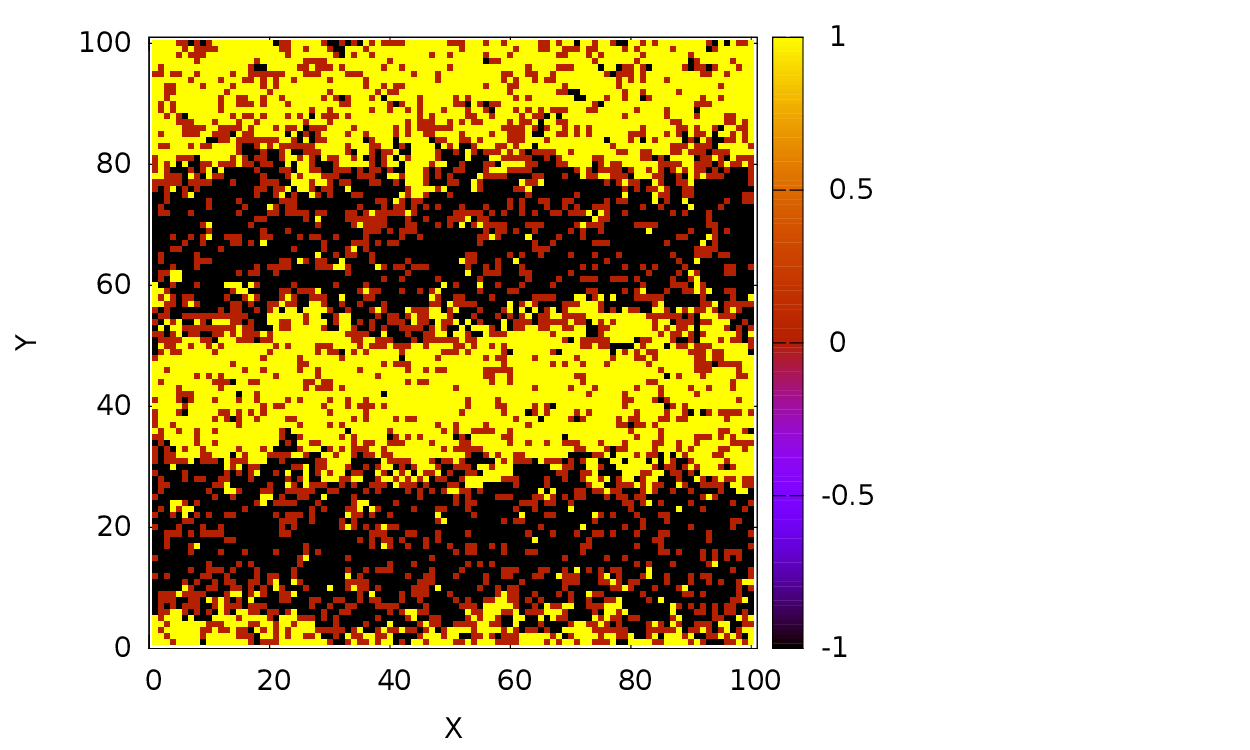}}
\resizebox{7cm}{!}{\includegraphics[angle=0]{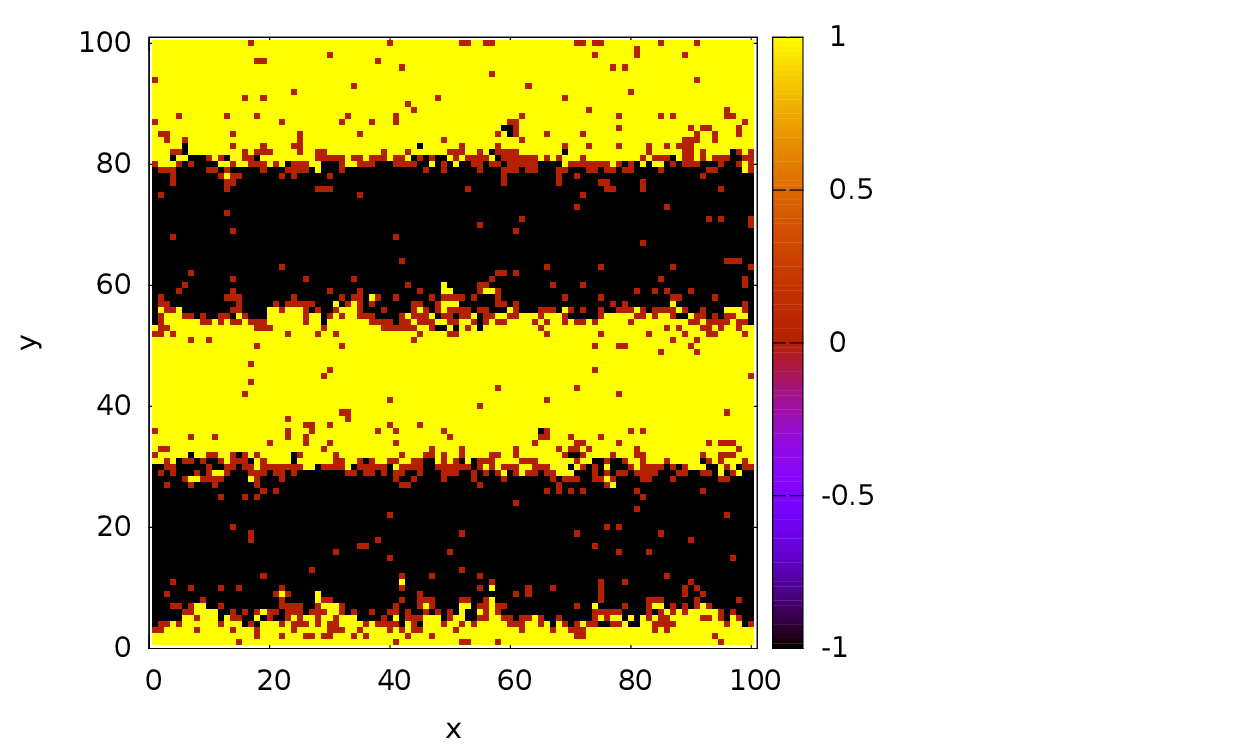}}
          \end{tabular}
\caption{Morphology of the lattice (value of $S_z(x,y,t=2000)$) for different values 
of T, H and D. Clockwise from top-left (i) D=1.0, H=0.5, T=0.5 (ii) D=0.1,
H=0.3, T=1.9, (iii) D=1.0, H=1.5, T=1.0 and (iv) D=1.0, H=0.5, T=1.5}

\end{center}
\end{figure}

\newpage
\begin{figure}[h]
\begin{center}
\begin{tabular}{c}
        \resizebox{7cm}{!}{\includegraphics[angle=0]{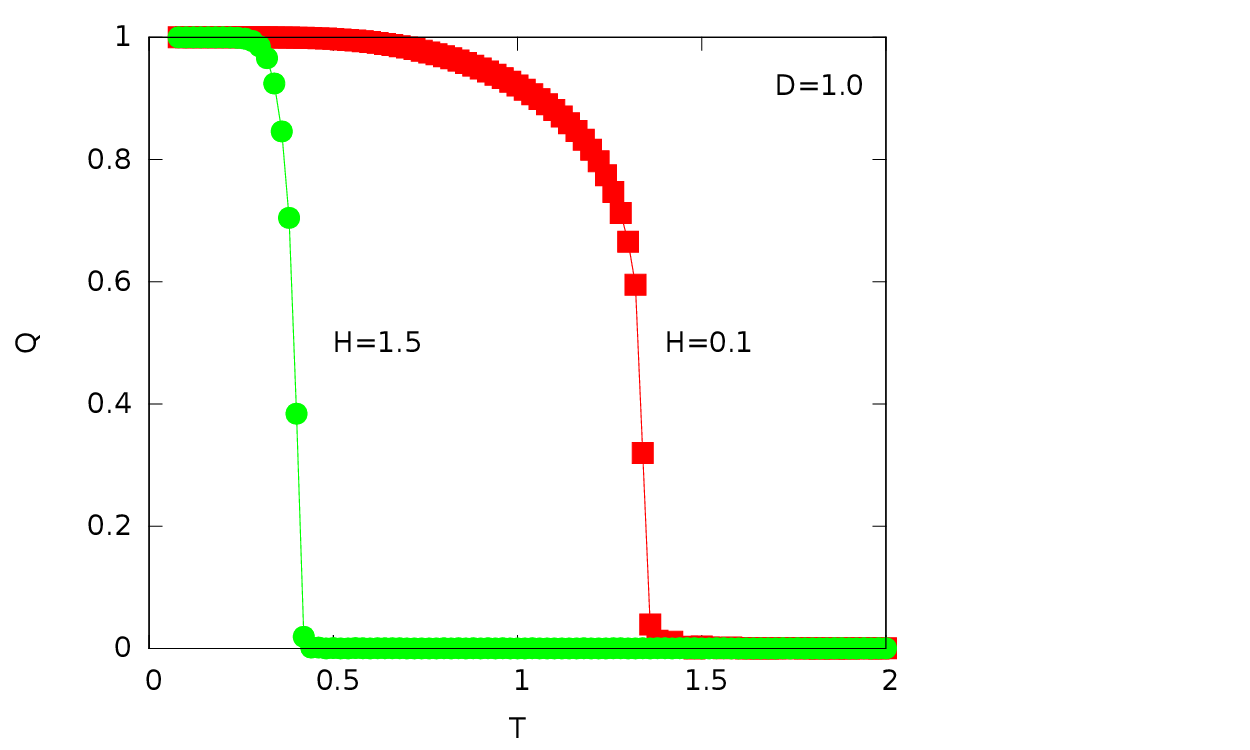}}
        \resizebox{7cm}{!}{\includegraphics[angle=0]{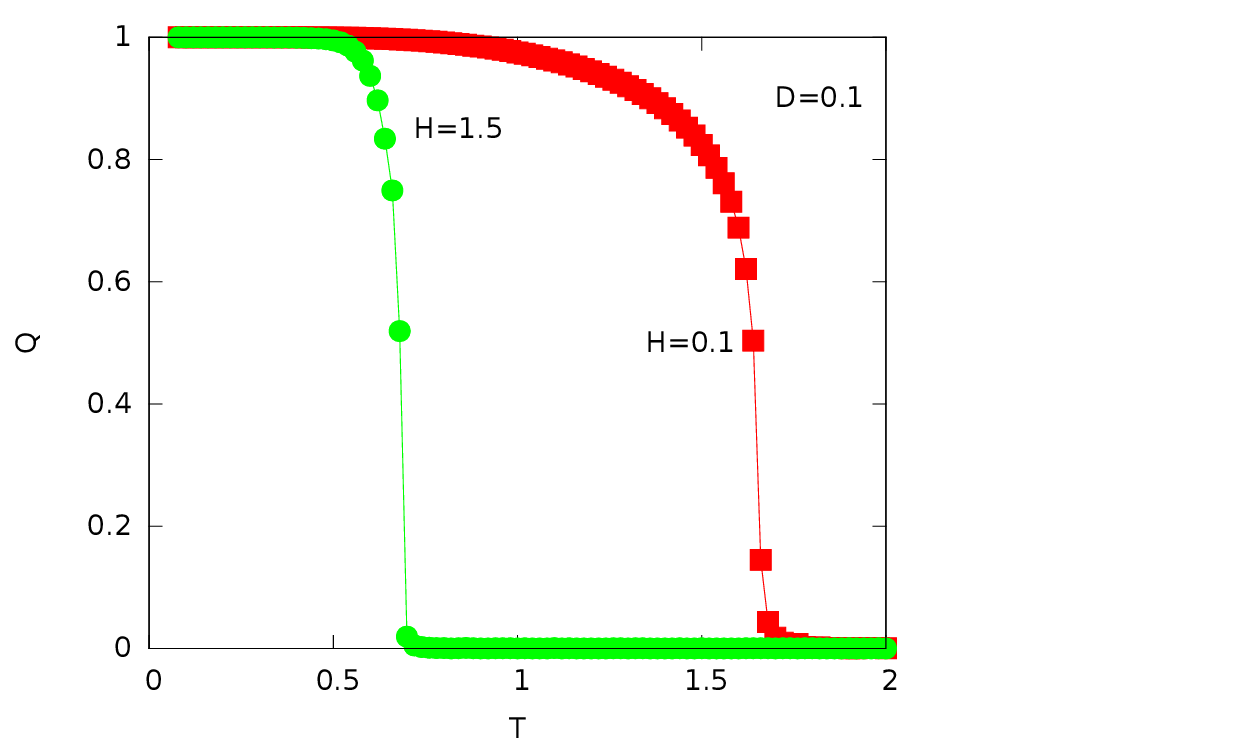}}
        \\
        \resizebox{7cm}{!}{\includegraphics[angle=0]{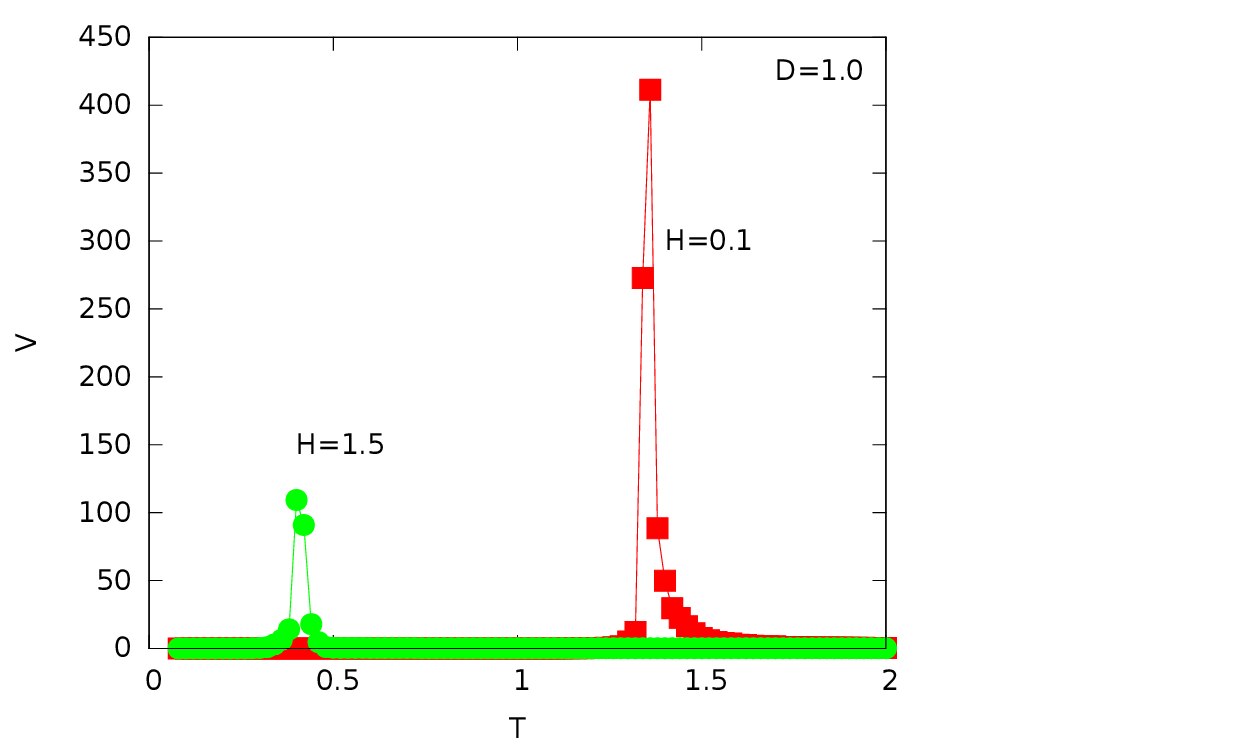}}
        \resizebox{7cm}{!}{\includegraphics[angle=0]{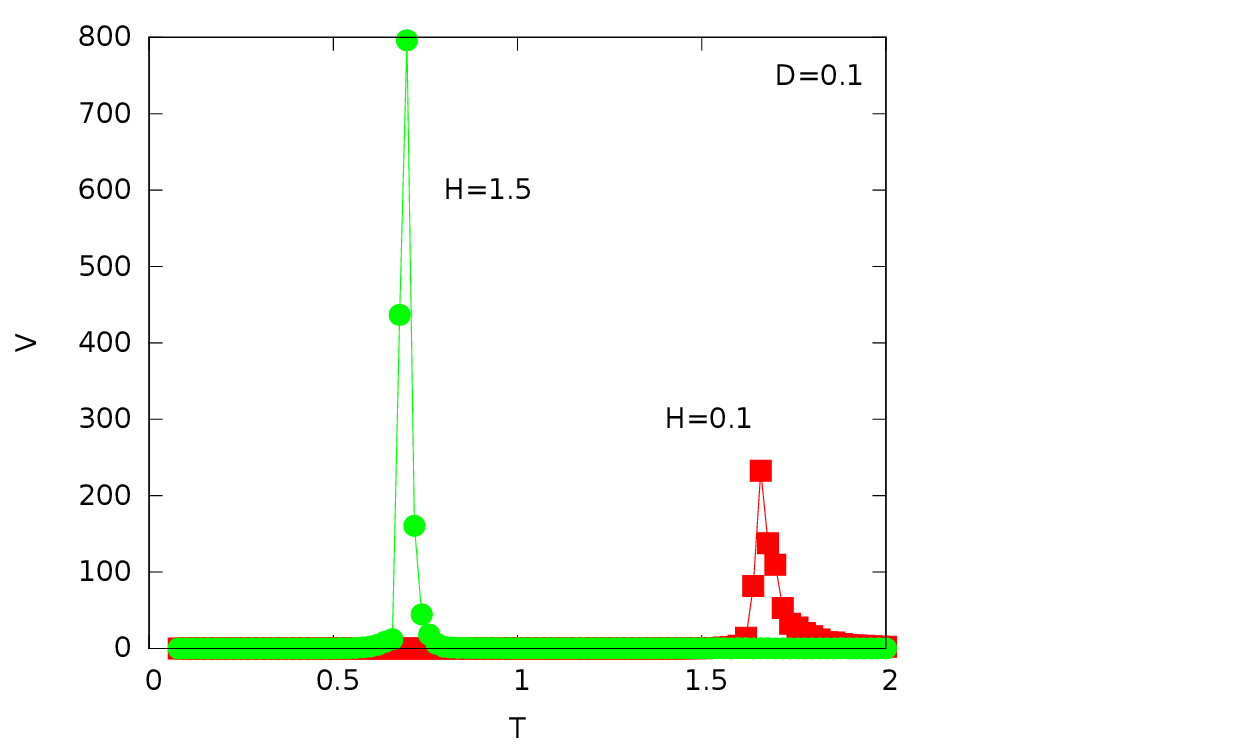}}
          \end{tabular}
 \caption{Temperature dependence of Q and V for two different values of H.
Here, D=1.0 (left) and D=0.1 (right).}
\end{center}
\end{figure}
\newpage
\begin{figure}[h]
\begin{center}
\begin{tabular}{c}
        \resizebox{10cm}{!}{\includegraphics[angle=0]{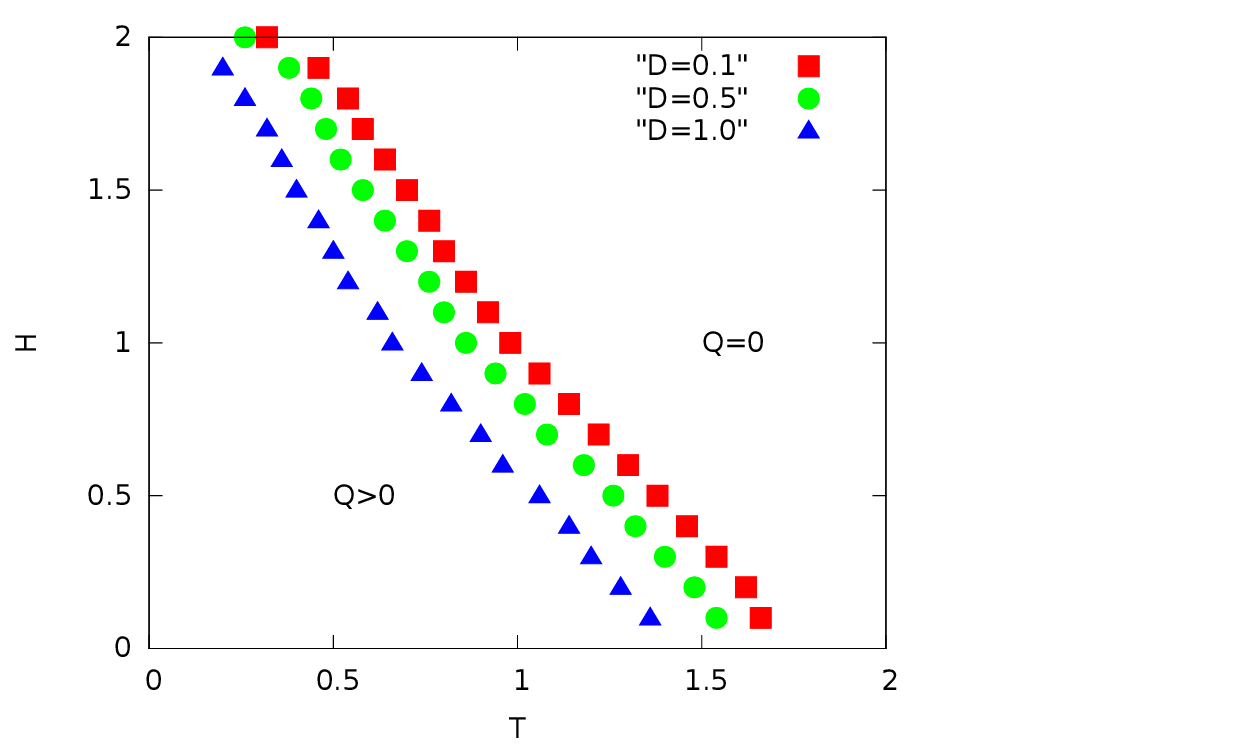}}
        
          \end{tabular}
 \caption{Phase diagram in the H-T plane. Different symbols denote different values
of D (mentioned in the figure).}
\end{center}
\end{figure}
\newpage
\begin{figure}[h]
\begin{center}
\begin{tabular}{c}
\resizebox{8cm}{!}{\includegraphics[angle=0]{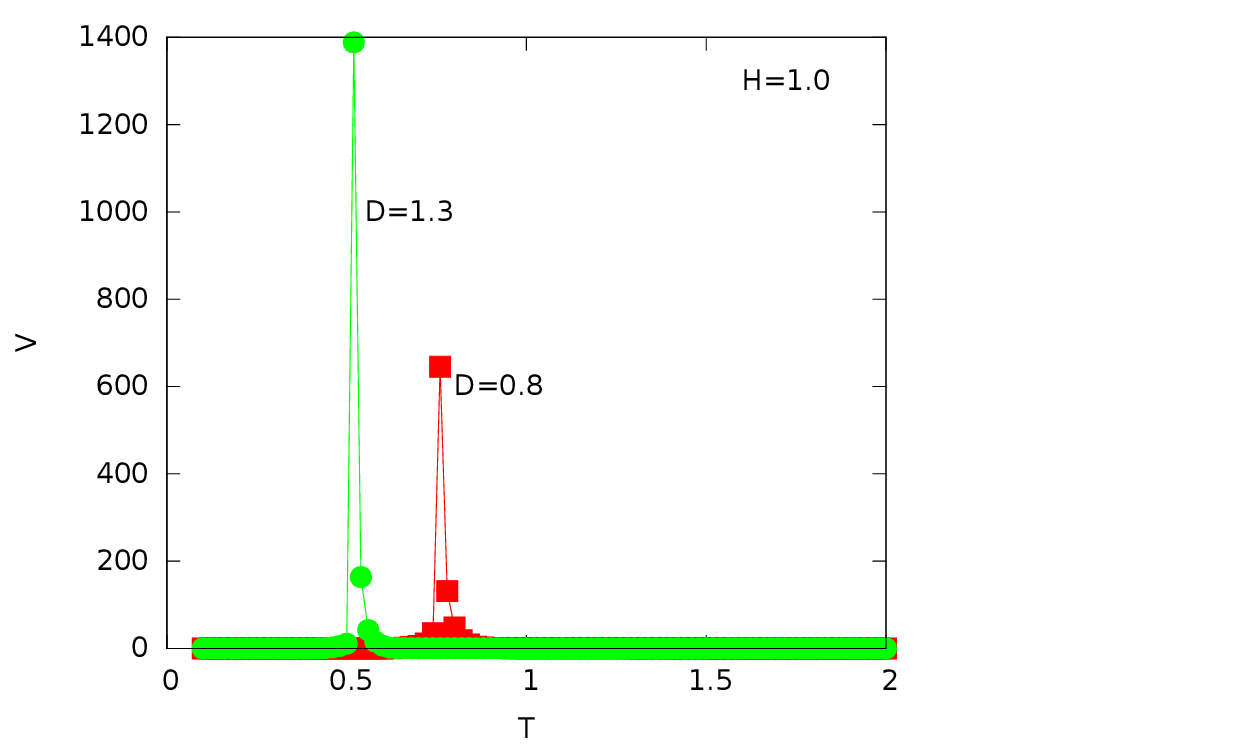}}
\\
\resizebox{8cm}{!}{\includegraphics[angle=0]{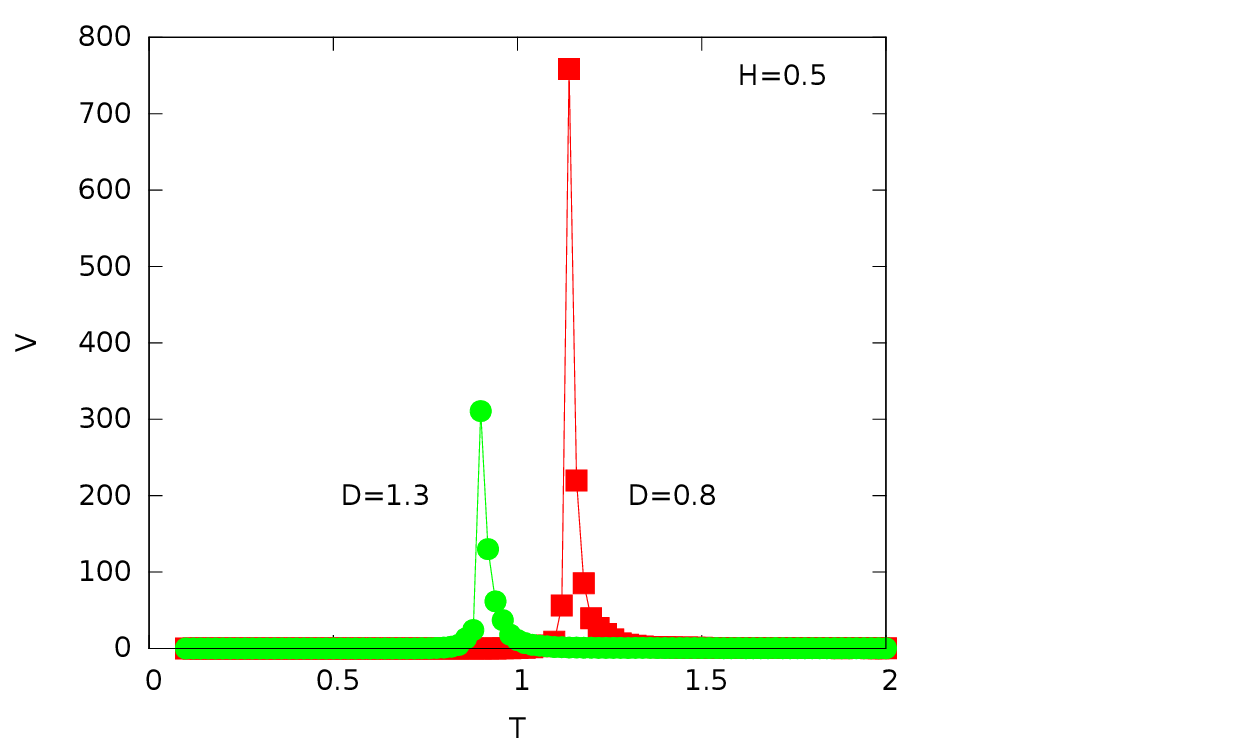}}
     
\end{tabular}
\caption{Temperature dependences of V for two different field amplitudes H and
D. D=0.8 (Red square) and D=1.3 (Green bullet).}
\end{center}
\end{figure}

\newpage
\begin{figure}[h]
\begin{center}
\begin{tabular}{c}
        \resizebox{10cm}{!}{\includegraphics[angle=0]{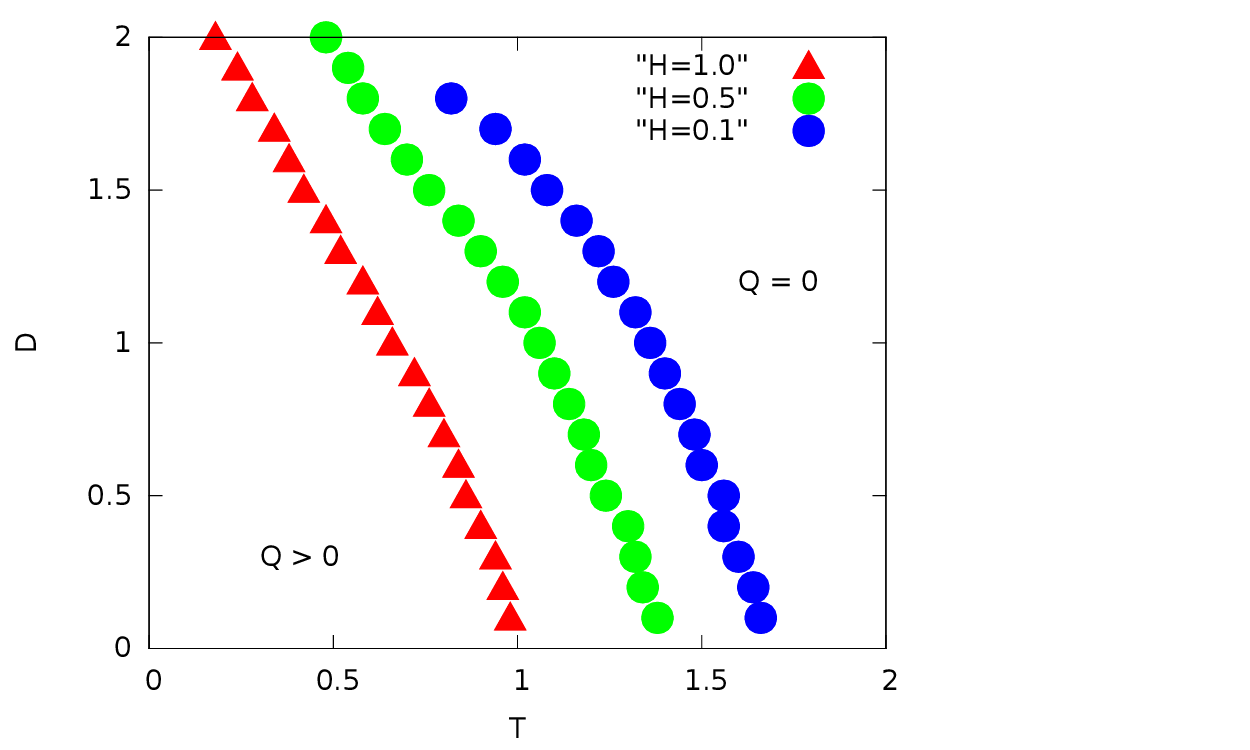}}
        
          \end{tabular}
 \caption{Phase diagram in the D-T plane in the case of propagating wave. Different symbols denote different values
of H (mentioned in the figure).}
\end{center}
\end{figure}
\newpage
\newpage
\begin{figure}[h]
\begin{center}
\begin{tabular}{c}
        \resizebox{8cm}{!}{\includegraphics[angle=0]{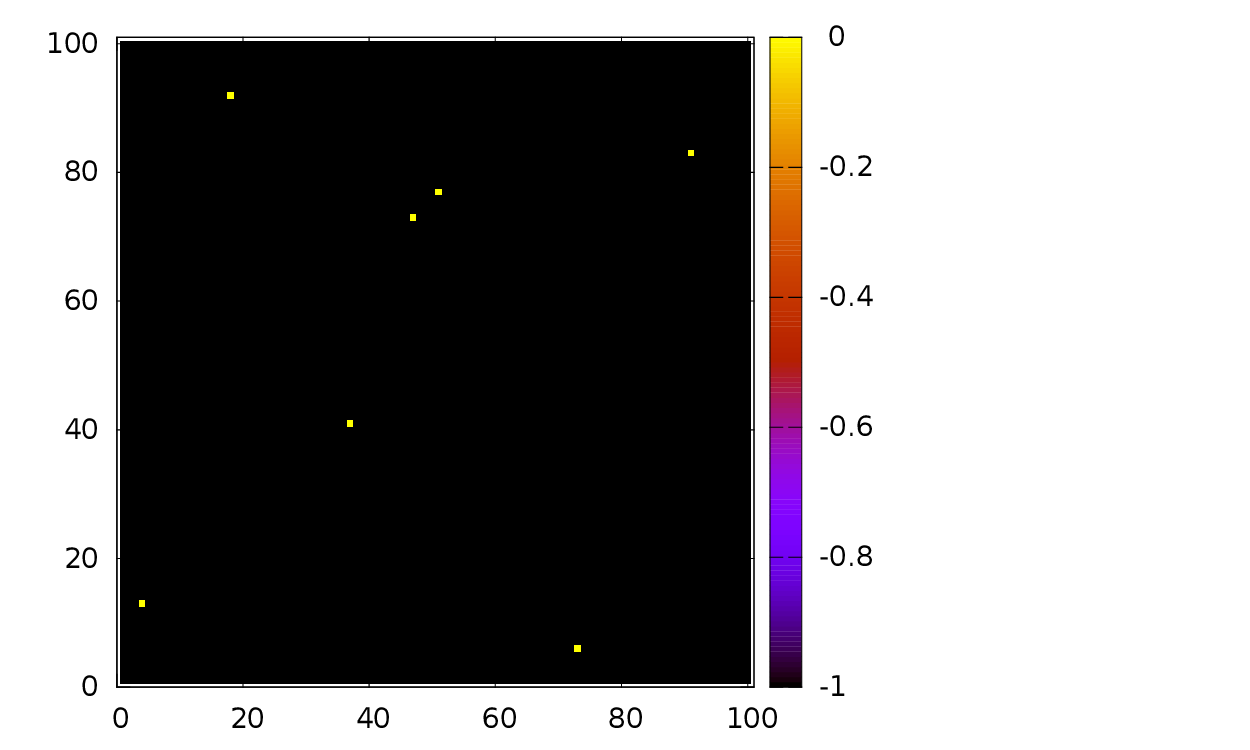}}
       \\        
        \resizebox{8cm}{!}{\includegraphics[angle=0]{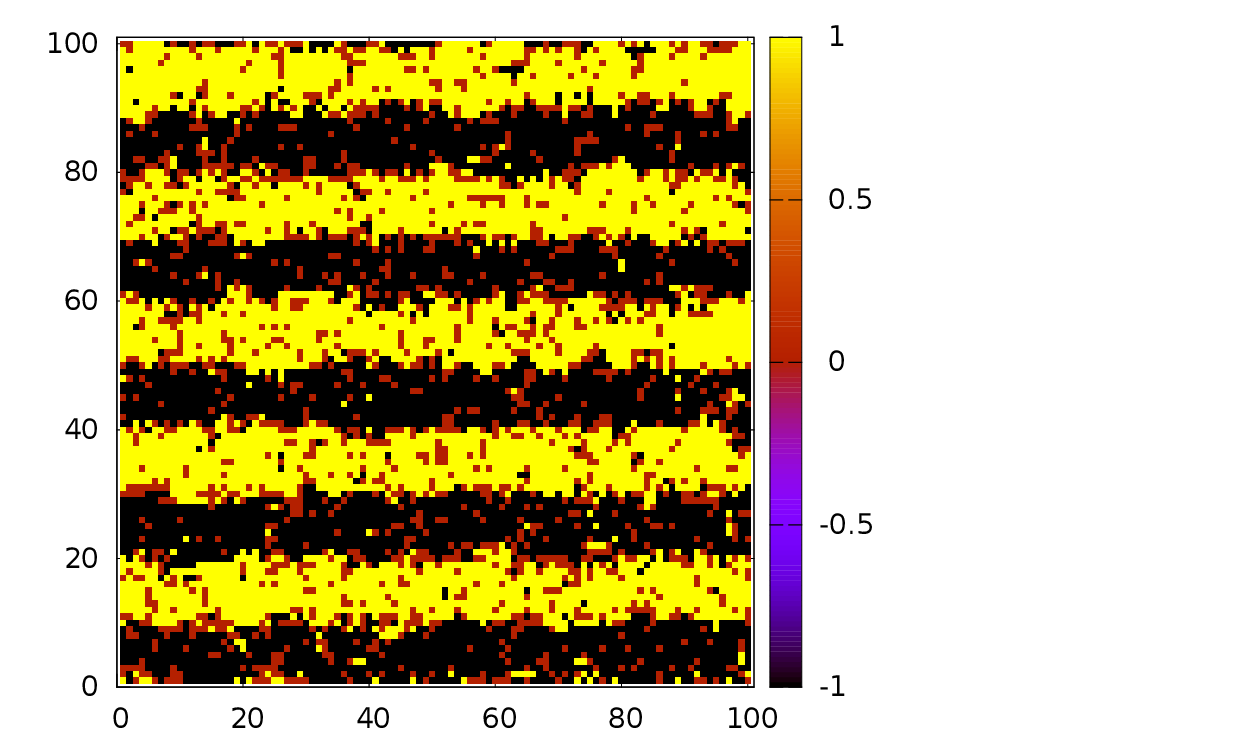}}
          \end{tabular}
 \caption{Morphologies of ordered or pinned 
(top for H=0.5, D=0.5 and T=0.5) and 
disordered or standing (bottom for H=1.5, D=0.5 and T=1.5) phases.
Here, $\lambda=20$.}
\end{center}
\end{figure}
\newpage
\begin{figure}[h]
\begin{center}
\begin{tabular}{c}
        \resizebox{8cm}{!}{\includegraphics[angle=0]{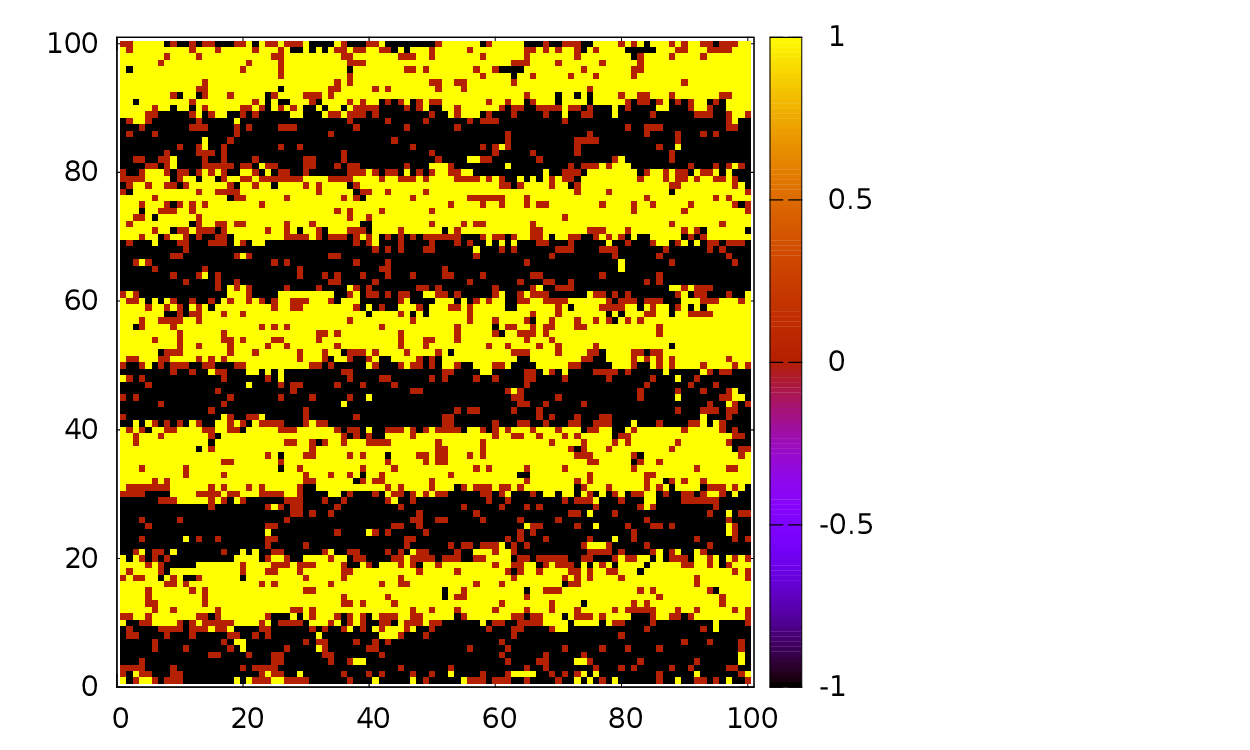}}
       \\        
        \resizebox{8cm}{!}{\includegraphics[angle=0]{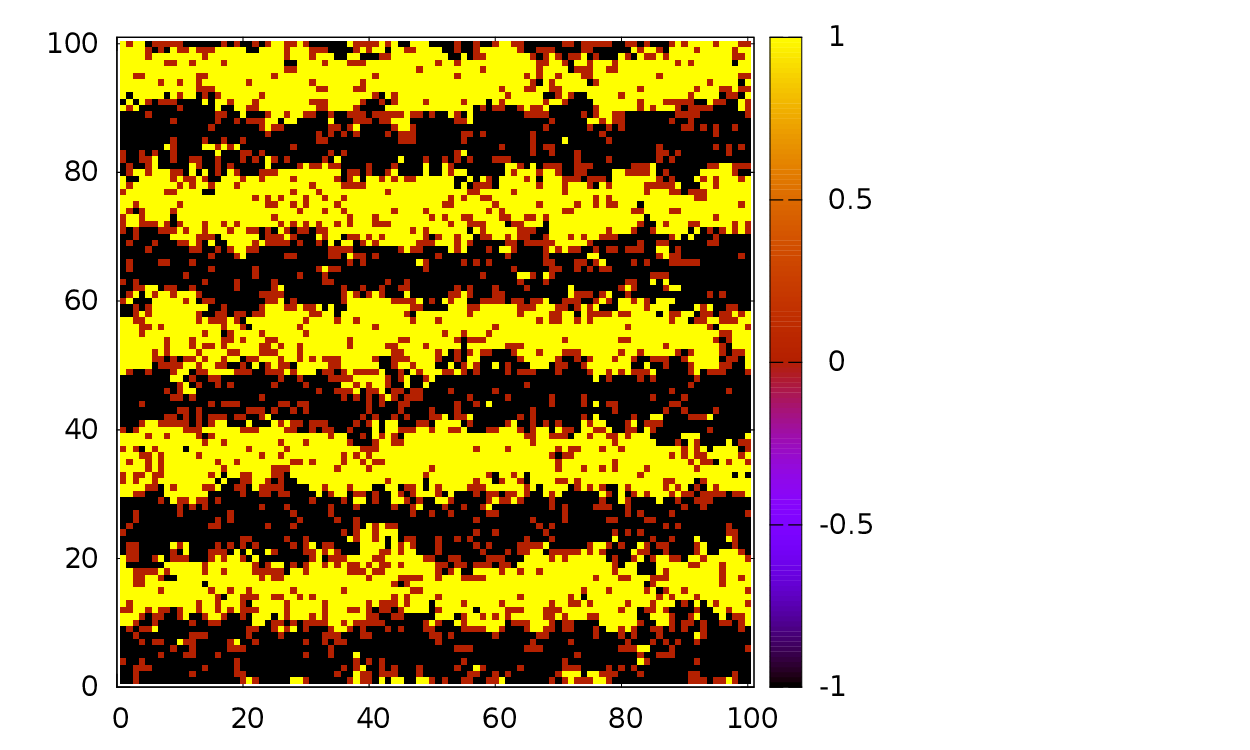}}
          \end{tabular}
 \caption{Morphologies of standing wave dynamical modes
(non-propagating) in disordered phase for two
different times. Top one is at $t=1000$ MCSS and the bottom one is taken
at $t=1070$ MCSS. Here, H=1.5, T=1.5 and D=0.5 $\lambda=20$.} 
\end{center}
\end{figure}
\newpage
\begin{figure}[h]
\begin{center}
\begin{tabular}{c}
        \resizebox{8cm}{!}{\includegraphics[angle=0]{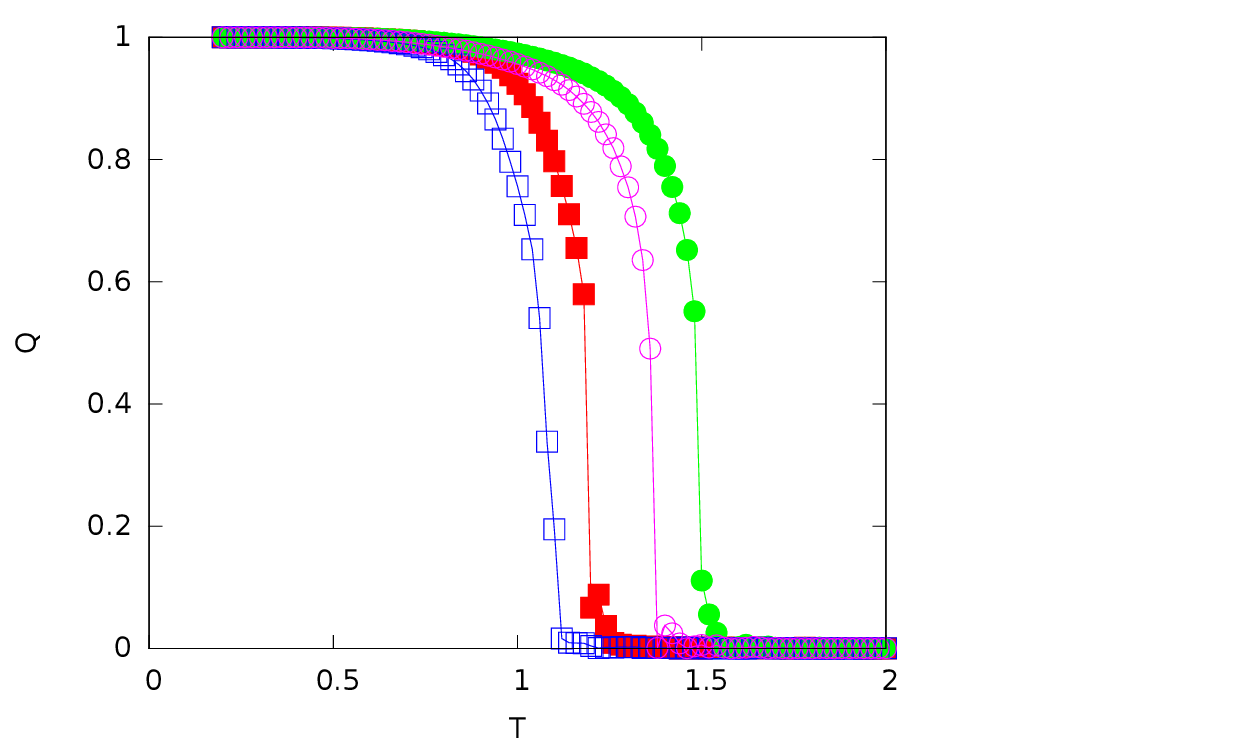}}
       \\ 
        \resizebox{8cm}{!}{\includegraphics[angle=0]{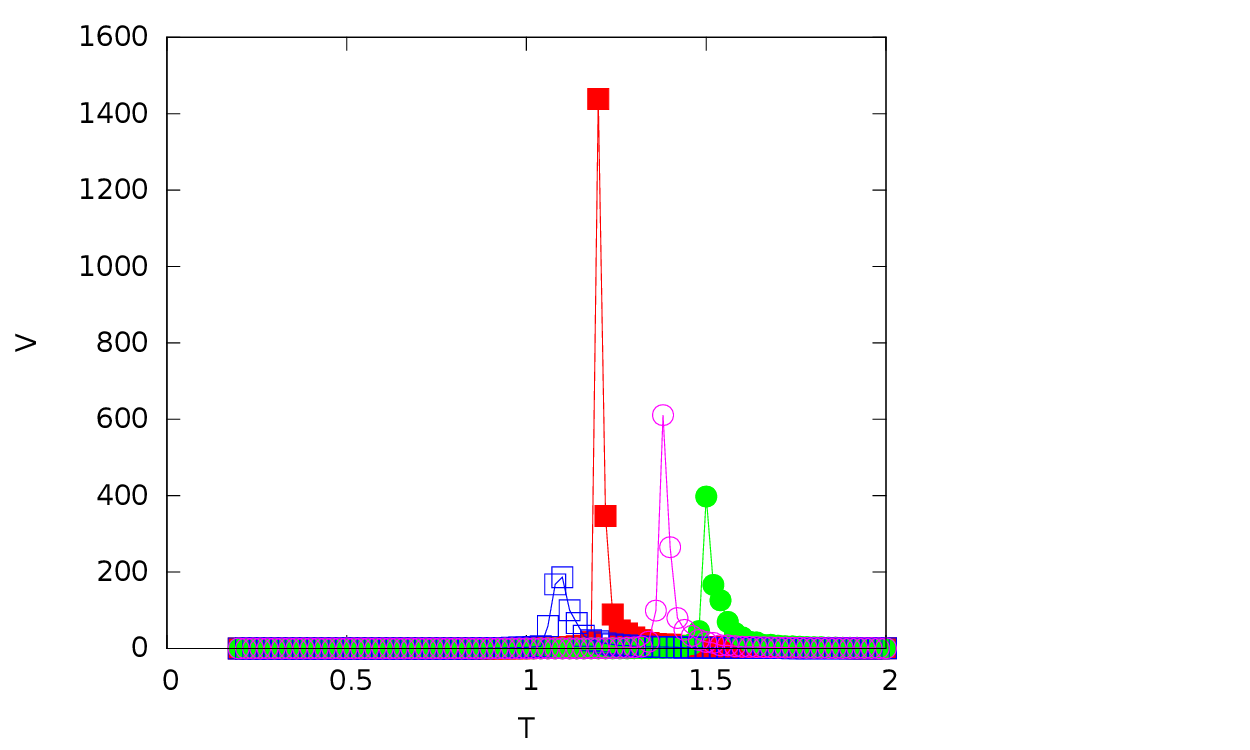}}
          \end{tabular}
 \caption{Q and V are plotted against temperature (T) for different values of
field amplitude (H) and anisotropy (D) in the case of standing wave of
wave length ($\lambda$) 20. Here, different symbols represent different values of H and D. (i) Red filled square (H=1.0, D=0.1) (ii) Blue open square (H=1.0,
D=0.5), (iii) Green filled circle (H=0.5, D=0.1) and (iv) Open magenta circle
(H=0.5, D=0.5)
}
\end{center}
\end{figure}
\newpage
\begin{figure}[h]
\begin{center}
\begin{tabular}{c}
        \resizebox{10cm}{!}{\includegraphics[angle=0]{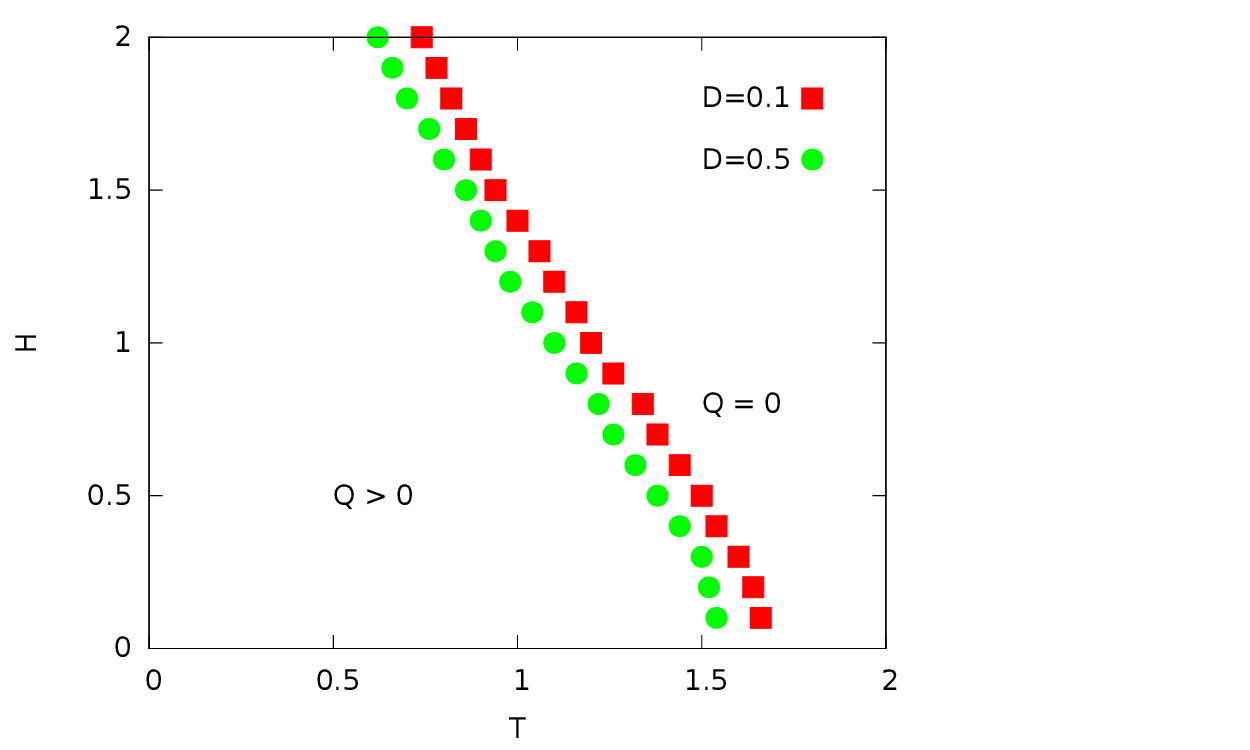}}
        
          \end{tabular}
 \caption{Phase diagram in the H-T plane in the case of standing wave. Different symbols denote different values
of D (mentioned in the figure). Here $\lambda=20$.}
\end{center}
\end{figure}
\newpage
\begin{figure}[h]
\begin{center}
\begin{tabular}{c}
        \resizebox{10cm}{!}{\includegraphics[angle=0]{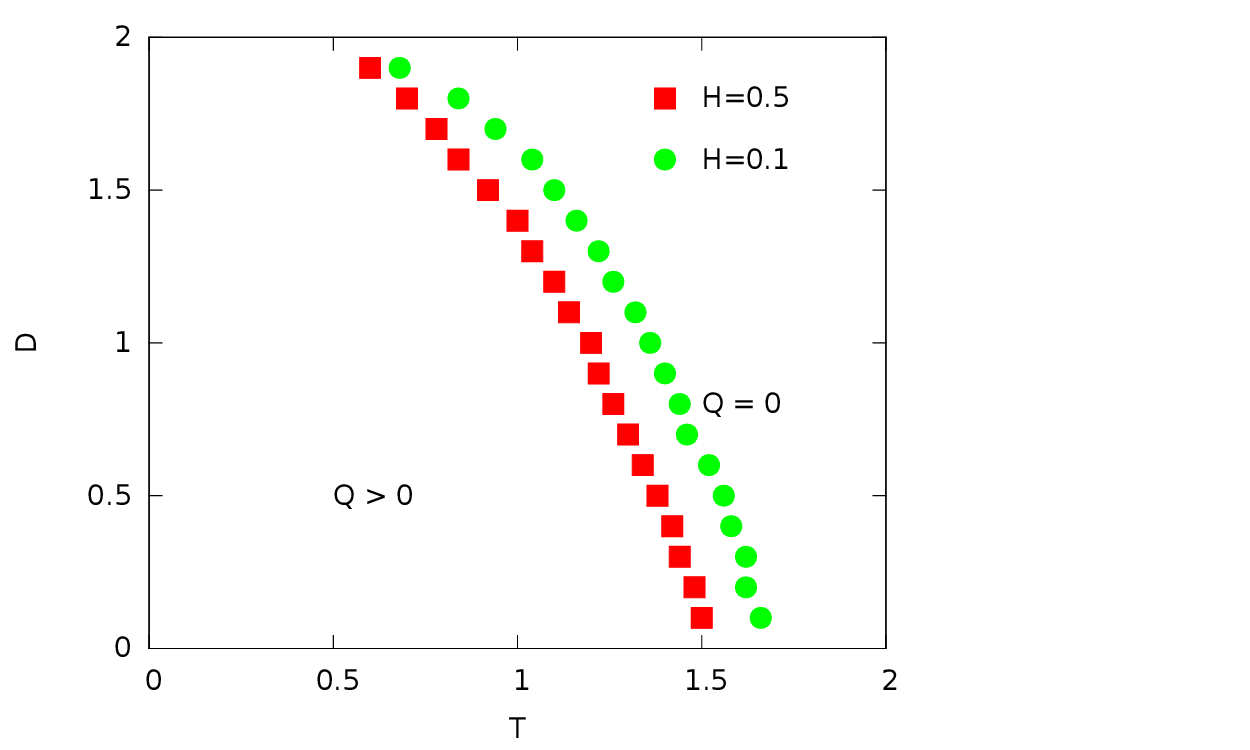}}
        
          \end{tabular}
 \caption{Phase diagram in the D-T plane in the case of standing wave. Different symbols denote different values
of H (mentioned in the figure). Here $\lambda=20$.}
\end{center}
\end{figure}
\newpage
\begin{figure}[h]
\begin{center}
\begin{tabular}{c}
        \resizebox{10cm}{!}{\includegraphics[angle=0]{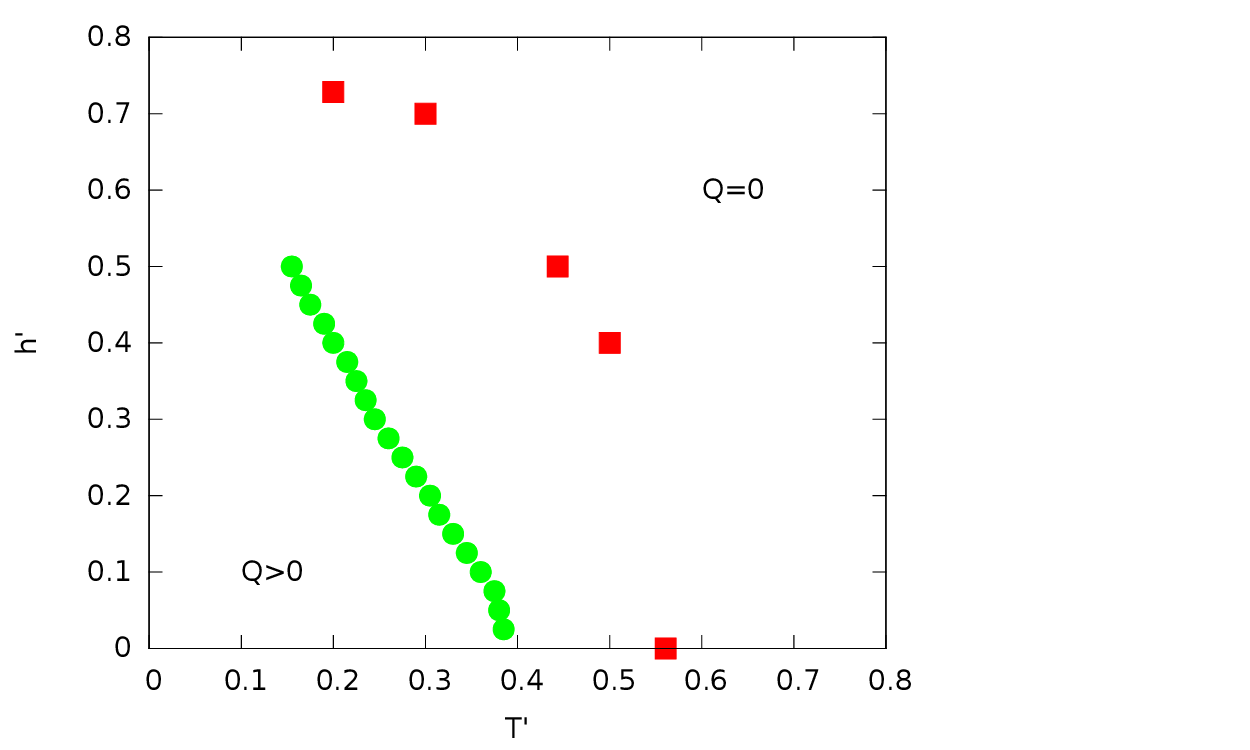}}
        
          \end{tabular}
 \caption{A typical comparison of phase diagrams. (i) (Red square) the Glauber 
kinetic $S=1$ Blume-Capel model swept by oscillating (uniform over space)
magnetic field in meanfield approximation. The data collected from Fig-7(b) of
the reference\cite{keskin}(ii) (Green bullet) The Monte Carlo 
metropolis results of $S=1$ BC model swept by standing magnetic wave. The 
$D=0.5$ and $\lambda=20$ here.}
\end{center}
\end{figure}
\end{document}